\documentclass[10pt]{iopart}

\usepackage{iopams}
\usepackage[dvips]{graphicx}
\usepackage{wrapfig}

\begin{document}

\title[Superposition of coherent states prepared in one mode...]{\Large{Superposition of coherent states prepared in one mode of a dissipative bimodal cavity}}

\author{Iara P. de Queir\'{o}s$^{1}$, W. B. Cardoso$^{1}$, and N. G. de Almeida$^{1,2,*}$}

\address{$^{1}$Instituto de F\'{\i}sica, Universidade Federal de Goi\'{a}s, 74.001-970, Goi%
\^{a}nia (GO) Brazil.}
\address{$^{2}$N\'{u}cleo de Pesquisas em F\'{\i}sica, Universidade Cat\'{o}lica de Goi\'{a}%
s, 74.605-220, Goi\^{a}nia (GO), Brazil.}

\ead{$^{*}$norton@pesquisador.cnpq.br}

\begin{abstract}
We solve the problem of the temporal evolution of one of two-modes
embedded in a same dissipative environment and investigate the role
of the losses after the preparation of a coherent state in only one
of the two modes. Based on current cavity QED technology, we present
a calculation of the \textit{fidelity} of a superposition of
coherent states engineered in a bimodal high-Q cavity. Our
calculation demonstrates that the engineered superposition retains
coherence for large times when the mean photon number of the
prepared mode is on the order of unity.

\end{abstract}


Problems involving interaction of atoms and two modes of a
electromagnetic field trapped into a high-Q cavity have increasingly
received attention [1-14]. From the theoretical point of view,
bimodal cavities were considered in a number of papers. For example,
in \cite{hafez92} the phenomena of a degenerate and nondegenerate
two-mode squeezing for a generalized
Jaynnes-Cummings model with a two-level atom was considered; in \cite%
{asoka04} the authors show how to prepare W states, GHZ states, and
two-qutrit entangled states using the multi-atom two-mode
entanglement; in \cite{geisa04,cardoso05,yang06} schemes for
teleporting entangled state of zero- and -one photon state from a
bimodal cavity to another were proposed; in
\cite{prado06,villasboas05} \ the authors show how to build bilinear
and quadratic Hamiltonians, thus opening the possibility to
implement up- and down-conversion operations in two mode cavity QED.
Recently, experiments with coherent control of atomic collision and
controlled entanglement in bimodal cavities have been reported
\cite{haroche01,osnaghi01}, attracting even more attention to this
topic.

It is to be noted that all the above mentioned papers do not take
into account the effect of the environment, which plays an important
role mainly when the decoherence time has to be considered. Damping
of two modes of a cavity QED has been considered in
Refs.\cite{gou89,napoli03,rossi04}, all of them using the Liouville
approach. In \cite{gou89} the behavior of \ a two-level atom
interacting with two modes of a light field in a cavity was
investigated, and the effects of cavity damping were treated
numerically for the special case of fields initially prepared in two
mode squeezed vacuum state. In ref. \cite{napoli03} the authors
studied the evolution of a two-level atom coupled to two modes of a
dissipative cavity, and in ref. \cite{rossi04} the authors have
investigated cross decay rates and robust (against dissipation)
states in a system composed by two cavity modes in the presence of
the same reservoir.

In this paper we review the problem of two modes in a lossy cavity
QED. Our model differs from that of \cite{napoli03} since we take
into account
bosonic modes in a dissipative environment. Also, different from \cite%
{gou89,rossi04}, our model includes, beside two non-interacting
bosonic modes subject to the same reservoir, a two-level atom
interacting off-resonantly with one of the two modes of the bimodal
cavity. Our
Hamiltonian model is%
\begin{equation}
H=H_{0}+H_{I}  \label{1}
\end{equation}%
where%
\begin{equation}
H_{0}=\sum\limits_{j=1}^{2}\hbar \omega _{j}a_{j}^{\dagger
}a_{j}+\sum\limits_{k}\hbar \omega _{k}b_{k}^{\dagger
}b_{k}+\frac{\hbar \omega _{0}}{2}\sigma _{z}+\hbar a_{1}^{\dagger
}a_{1}\chi _{1}\sigma _{z} ,  \label{2a}
\end{equation}%
and%
\begin{equation}
H_{I}=\sum\limits_{k}\sum\limits_{j=1}^{2}\hbar \left( \lambda
_{jk}a_{j}^{\dagger }b_{k}+\lambda _{jk}^{\ast }a_{j}b_{k}^{\dagger
}\right) .  \label{3a}
\end{equation}%
Here we consider a two-level atom composed by ground $\left\vert
g\right\rangle $ and excited $\left\vert e\right\rangle $ state, $\
\sigma _{z}=\left\vert e\right\rangle \left\langle e\right\vert
-\left\vert g\right\rangle \left\langle g\right\vert ,$
$a_{j}^{\dagger }$ and $a_{j}$ are, respectively, the creation and
annihilation operator for the j$th$ cavity mode of frequency $\omega
_{j}$, whereas $b_{k}^{\dagger }$ and $b_{k} $ are the analogous
operators for the k$th$ reservoir oscillator mode, whose
corresponding frequency and coupling with the mode $j=1,2$ write
$\omega _{k}
$ and $\lambda _{jk}$, respectively. The atom-field coupling parameter is $%
\chi _{1}=\frac{g^{2}}{\delta _{1}}$, where $g$ is the Rabi frequency and $%
\delta _{1}=\left( \omega _{1}-\omega _{0}\right) $ is the detuning
between the field frequency $\omega _{1}$\ and the atomic frequency
$\omega _{0}$. Note that the dispersive interaction occurs with only
mode $1$. This is important for preparing superposition of coherent
states \cite{brune96}. Using the completeness relation for both atom
and fields as given by coherent states, the Schr\"{o}dinger state
vector associated with Hamiltonian Eq.(\ref{1}) can be written as
\begin{equation}
\left\vert \Phi (t)\right\rangle =\left\vert g\right\rangle
\left\vert \phi _{g}(t)\right\rangle +\left\vert e\right\rangle
\left\vert \phi _{e}(t)\right\rangle   \label{4a}
\end{equation}%
where $\left\vert \phi _{\ell }(t)\right\rangle =\int \frac{d^{2}\alpha _{1}%
}{\pi }\int \frac{d^{2}\alpha _{2}}{\pi }\int \left\{ \frac{d^{2}\beta _{k}}{%
\pi }\right\} \mathcal{A}_{\ell }\left\vert \alpha _{1},\alpha
_{2},\left\{
\beta _{k}\right\} \right\rangle $, $\ell =g,e$. The complex quantities $%
\alpha _{1},\alpha _{2},\left\{ \beta _{k}\right\} $ stand for the
eigenvalues of $a_{j}$ and $b_{k}$, respectively, and
$\mathcal{A}_{\ell }\left( \alpha _{1},\alpha _{2},\left\{ \beta
_{k}\right\} ,t\right) $ are the expansion coefficients for
$\left\vert \phi _{\ell }(t)\right\rangle $ in the basis of the
coherent state products $\left\vert \alpha _{1},\alpha
_{2},\left\{ \beta _{k}\right\} \right\rangle $. The assumption of $%
\left\vert \left\{ \beta _{k}\right\} \right\rangle $\ for the
reservoir
follows from a remarkable property first proved by Mollow and Glauber \cite%
{glauber67}, \textit{i.e}., that at zero absolute temperature the
reservoir receives coherently the excitation lost by a coherent
state. The extension to finite temperature is considered in
Ref.\cite{norton00}, where a phenomenological-operator approach to
dissipation in cavity QED was proposed. Essentially, this extension
consists in treating the reservoir \textit{plus} the surrounding as
a pure state provided that we include the degrees of freedom of
whatever remain\ in the surrounding, in such way that we obtain a
mixed (thermal) state after tracing out the degrees of freedom of
whatever was included. Because the orthogonality of the atomic
states and
Eq.(\ref{1}-\ref{4a}), we can obtain the uncoupled time-dependent Schr\"{o}%
edinger equations%
\begin{equation}
i\hslash \frac{d}{dt}\left\vert \phi _{\ell }(t)\right\rangle
=H_{\ell }\left\vert \phi _{\ell }(t)\right\rangle ,  \label{5a}
\end{equation}%
where $H_{\ell }=H_{\ell 0}+H_{I}$, and%
\begin{equation}
H_{\ell 0}=\hbar \left( \omega _{1}\pm \chi \right) a_{1}^{\dagger
}a_{1}+\hbar \omega _{2}a_{2}^{\dagger }a_{2}+\sum\limits_{k}\hbar
\omega _{k}b_{k}^{\dagger }b_{k}.  \label{4}
\end{equation}%
Note that the problem has been reduced to that of dissipation of two
modes of the cavity field whose frequency $\omega _{1}$ of mode $1$\
has been shifted by $-\chi $ ($\chi $) when interacting with the
ground (excited) state of the atom. To solve the problem of \ the
evolution of an arbitrary initial state prepared in one of the two
modes we pursue a different approach from that of Refs.
\cite{gou89,napoli03,rossi04}. Instead of following the evolution of
the two modes at once, we will follow rightly to the mode of
interest, assumed as mode $1$. A convenient way to perform this can
be done by the method of the reduced density operator \cite{mandel}.
For this purpose we first calculate the characteristic function $G$\
in the normal order, then the Glauber-Sudarshan $P$ representation,
and finally the reduced density operator $\rho _{1}(t)$ for mode
$1$. The $G$ function for
mode $1$ in the Heisenberg picture reads%
\begin{equation}
G(\eta ,\eta ^{\ast },t)=Tr_{2R}\left\{ \rho _{12R}(0)\exp \left[
\eta a_{1}^{\dagger }(t)\right] \exp \left[ -\eta ^{\ast
}a_{1}(t)\right] \right\} ,~~~~  \label{5aa}
\end{equation}%
where $\rho _{12R}(0)$ is the density operator for the whole system
composed by modes $1$,$2$ and the reservoir at the instant $t=0$,
and $Tr_{2R}$ indicates the trace on mode 2 and reservoir. The
Glauber-Sudarshan $P$ representation is given by the two-dimensional
Fourier transform of the
characteristic function $G$:%
\begin{equation}
P\left( \xi ,\xi ^{\ast },t\right) =\int G(\eta ,\eta ^{\ast
},t)\exp \left( -\eta \xi ^{\ast }+\eta ^{\ast }\xi \right)
\frac{d^{2}\eta }{\pi ^{2}} ,  \label{5bb}
\end{equation}%
while the reduced density operator for mode $1$ of the cavity field
state is
given by the diagonal representation of $P\left( \xi ,\xi ^{\ast },t\right) $:
\begin{equation}
\rho _{1}(t)=\int P\left( \xi ,\xi ^{\ast },t\right) \left\vert \xi
\right\rangle \left\langle \xi \right\vert d^{2}\xi . \label{5cc}
\end{equation}%
To apply this method, we will need to solve the following Heisenberg
equations for modes $a_{j}$ corresponding to the Hamiltonian
Eq.(\ref{4}) in order to substitute them in Eq.(\ref{5aa}):
\begin{eqnarray}
\dot{a_{j}} &=&-i\omega _{j\ell }a_{j}-i\sum\limits_{k}\lambda
_{ik}^{\ast }b_{k}  \label{5} \\
\dot{b_{k}} &=&-i\omega _{k}b_{i}-i\sum\limits_{i=1}^{2}\lambda
_{ik}a_{i}, \label{6}
\end{eqnarray}%
where $\omega _{j\ell }=\left( \omega _{1}\pm \chi \right) $ if
$j=1$ and the minus (plus) signal depends on the initial state $g$
($e$) of the atom, or $\omega _{j\ell }=\omega _{2}$ if $j=2$.
Eqs.(\ref{5}-\ref{6}) can be solved, for example, by Laplace
transform. In doing so, we multiply both
sides of Eqs.(\ref{5}-\ref{6}) by $\exp (-st),$ integrate from $0$ to $%
\infty $ and evaluate the inverse Laplace transform under the
Wigner-Weisskopf approximation \cite{mandel}%
\begin{equation*}
\sum_{k}\frac{\lambda _{kj}^{\ast }\lambda _{kj^{\prime }}}{s+i\omega _{k}}%
\rightarrow i\Delta \omega _{jj^{\prime }}+\frac{\gamma _{jj^{\prime
}}}{2};~~~~~~j,~j^{\prime }=1,2
\end{equation*}%
where $\Delta \omega _{jj}$, $\gamma _{jj}$ denote the shift in the
energy and the decay rate of the $j$-mode, respectively, and $\Delta
\omega _{jj^{\prime }},$ $\gamma _{jj^{\prime }}$ ($j\neq j^{\prime
}$) the
corresponding cross shift in the energy and the cross decay rates \cite%
{rossi04} for modes $1$,$2$. The results for modes $1$ and $2$ are%
\begin{eqnarray}
a_{1}(t)
&=&u_{11}(t)a_{1}(0)+u_{12}(t)a_{2}(0)+\sum\limits_{k}\vartheta
_{1k}(t)b_{k}(0),~~~~~~~  \label{7a} \\
a_{2}(t)
&=&u_{22}(t)a_{2}(0)+u_{21}(t)a_{1}(0)+\sum\limits_{k}\vartheta
_{2k}(t)b_{k}(0),~~~~~~~  \label{8a}
\end{eqnarray}%
where
\begin{eqnarray}
\hspace{-2.5cm}u_{11}(t)&=&\exp\left[ -\frac{(A+B)}{2}t\right] \left[
\frac{\left( B-A\right) }{\sqrt{\left( B-A\right) ^{2}+4\left\vert
C\right\vert ^{2}}} \sinh \left( \sqrt{\frac{\left( B-A\right) ^{2}+4\left\vert C\right\vert ^{2} }{2}}t\right) \right. \nonumber \\ \hspace{-2.5cm}&+& \left.  \cosh \left(
\sqrt{\frac{\left( B-A\right) ^{2}+4\left\vert
C\right\vert ^{2}}{2}}t\right) \right] \nonumber \\ \hspace{-2.5cm}\label{9a} \\
\hspace{-2.5cm}u_{12}(t) &=& - \exp \left[ -\frac{(A+B)}{2}t\right] \left[ \frac{2C}{\sqrt{%
\left( B-A\right) ^{2}+4\left\vert C\right\vert ^{2}}} \sinh \left( \sqrt{%
\frac{\left( B-A\right) ^{2}+4\left\vert C\right\vert ^{2}}{2}}t\right) %
\right], \nonumber \\ \hspace{-2.5cm}\label{10a}
\end{eqnarray}%
and
\begin{eqnarray}
\hspace{-2.5cm}\vartheta _{1k}(t) &=&i(\lambda _{k2}^{\ast }C-\lambda _{k1}^{\ast
}-\lambda
_{k1}^{\ast }B) \nonumber \\ \hspace{-2.5cm}&\times& \left\{ \frac{\exp \left[ -\left( \frac{A+B-\sqrt{%
(B-A)^{2}+4\left\vert C\right\vert ^{2})}}{2}\right) t \right] }{(\frac{%
-(B+A)+\sqrt{(B-A)^{2}+4\left\vert C\right\vert ^{2}}}{2}+i\omega _{k})\sqrt{%
(B-A)^{2}+4\left\vert C\right\vert ^{2}}}\right.  \nonumber \\
\hspace{-2.5cm}&+&\frac{\exp \left[ -(\frac{A+B+\sqrt{(B-A)^{2}+4\left\vert
C\right\vert ^{2})}}{2}) t \right]
}{(\frac{(B+A)+\sqrt{(B-A)^{2}+4\left\vert C\right\vert
^{2}}}{2}-i\omega _{k})\sqrt{(B-A)^{2}+4\left\vert C\right\vert
^{2}}} \nonumber \\
\hspace{-2.5cm}&-&\left. \frac{\exp (-i\omega
t)}{(\frac{(B+A)+\sqrt{(B-A)^{2}+4\left\vert C\right\vert
^{2}}}{2}-i\omega _{k})(\frac{-(B+A)+\sqrt{(B-A)^{2}+4\left\vert
C\right\vert ^{2}}}{2}+i\omega _{k})}\right\}
\end{eqnarray}%
with
\begin{eqnarray}
A &=&i\left( \omega _{1}\pm \chi +\Delta \omega _{1}\right) +\gamma
_{11}/2
\label{11a} \\
B &=&i\left( \omega _{2}\pm \chi +\Delta \omega _{2}\right) +\gamma
_{12}/2
\label{12a} \\
C &=&-i\Delta \omega _{12}-\gamma _{12}/2.  \label{14a}
\end{eqnarray}%
Functions$\ u_{21}(t)$, $u_{22}(t)$ and $\vartheta _{2k}(t)$ can be
obtained from $u_{12}(t)$, $u_{11}(t)$ and $\vartheta _{1k}(t)$,
respectively, simply replacing $C$ by $C^{\ast }$, $A$ by $B$, $1$
by $2$ and \textit{vice-versa}.

It is now straightforward to obtain $\rho _{1}(t)$ for an arbitrary
initial state if we note that any initial state
$\sum_{n,m}C_{nm}\left\vert m\right\rangle _{1}\left\vert
n\right\rangle _{2}$ can be 1written in its
most general form as%
\begin{eqnarray}
\rho _{12R}(0)&=&\sum\limits_{n,m,j,k}\int d^{2}\alpha _{1}d^{2}\alpha
_{2}d^{2}\beta _{1}d^{2}\beta _{2}C_{nmjk}\left\vert \alpha
_{1}\right\rangle _{11}\left\langle \alpha _{2}\right\vert
\left\vert \beta _{1}\right\rangle _{22}\left\langle \beta
_{2}\right\vert \nonumber \\ &\times& \int \left\{ d^{2}\gamma _{k}\right\} P\left(
\left\{ \gamma _{k}\right\} \right) \left\vert \left\{ \gamma
_{k}\right\} \right\rangle _{R}\left\langle \left\{ \gamma
_{k}\right\} \right\vert  \label{16a}
\end{eqnarray}%
where modes $1$ and $2$ were expanded in coherent states and the
reservoir is in a thermal state $\sum\limits_{k}p_{k}\left\vert
n_{k}\right\rangle _{R}\left\langle n_{k}\right\vert $ characterized
by an average photon number in the k$th$ mode $\left\langle
n_{k}\right\rangle =1/\left[ \exp (\hslash \omega _{k}/kT-1)\right]
$\ according to $p_{k}=\frac{\left\langle n_{k}\right\rangle
^{n}}{\left( 1+n_{k}\right) ^{n+1}}$, and was expanded in the
diagonal representation with $P\left( \left\{ \gamma _{k}\right\}
\right) =$ $\prod\limits_{k}\frac{1}{\pi \left\langle n_{k}\right\rangle }%
\exp \left( -\frac{\left\vert \gamma _{k}\right\vert
^{2}}{\left\langle n_{k}\right\rangle }\right) $. The characteristic
function for mode $1$
corresponding to the initial state of Eq.(\ref{16a}) reads%
\begin{eqnarray}
\hspace{-2.5cm}G(\eta ,\eta ^{\ast },t) &=&\sum\limits_{n,m,j,k}\int d^{2}\alpha
_{1}d^{2}\alpha _{2}d^{2}\beta _{1}d^{2}\beta _{2}C_{nmjk~~%
2}\left\langle \beta _{2}\right\vert \left. \beta _{1}\right\rangle
_{21}\left\langle \alpha _{2}\right\vert \left. \alpha
_{1}\right\rangle _{1} \nonumber \\ \hspace{-2.5cm}&\times& \exp \left[ -\eta ^{\ast }u_{11}(t)\alpha _{01}+\eta u_{11}^{\ast }(t)\alpha _{02}^{\ast }\right] \exp \left[ -\eta ^{\ast }u_{12}(t)\beta _{01}+\eta
u_{12}^{\ast }(t)\beta _{02}^{\ast }\right] \nonumber \\ \hspace{-2.5cm}&\times& \prod\limits_{k}\exp
\left[ -\left\vert \eta \right\vert ^{2}\left\vert \vartheta
_{1k}(t)\right\vert ^{2}\left\langle n_{k}\right\rangle \right] ,
\label{16}
\end{eqnarray}%
where we have written $f(0)=f_{0}$. The Glauber-Sudarshan $P\left(
\xi ,\xi ^{\ast },t\right) $ representation for mode $1$ as given by
Eq.(\ref{5bb})
is the Fourier transform of Eq.(\ref{16}):%
\begin{eqnarray}
\hspace{-2.5cm}
P\left( \xi ,\xi ^{\ast },t\right)&=&\sum\limits_{n,m,j,k}\int
d^{2}\alpha _{1}d^{2}\alpha _{2}d^{2}\beta _{1}d^{2}\beta
_{2}C_{nmjk}(\beta \alpha )\frac{_{1}\left\langle \alpha
_{2}\right\vert \left. \alpha _{1}\right\rangle _{12}\left\langle
\beta _{2}\right\vert \left. \beta
_{1}\right\rangle _{2}}{\pi D(t)} \nonumber \\ \hspace{-2.5cm}
&\times&\exp \left( -\frac{\left[ \xi -u_{11}(t)\alpha _{01}-u_{12}(t)\beta _{10}%
\right] \left[ \xi ^{\ast }-u_{11}^{\ast }(t)\alpha _{02}^{\ast
}-u_{12}^{\ast }(t)\beta _{20}^{\ast }\right] }{D(t)}\right) \label{17}
\end{eqnarray}%
where $D(t)=\sum \left\vert \vartheta _{1k}(t)\right\vert
^{2}\left\langle n_{k}\right\rangle $. The reduced density operator
for mode $1$ can now be obtained using Eq.(\ref{5cc}).

In the following, let us consider the important case of engineering
two coherent states inside a single bimodal high$-Q$ cavity with one
of the two modes supporting a coherent state and the other mode
being prepared in a superposition of coherent state (SCS) in the
presence of a reservoir cooled to $0$ K, \textit{i.e.}, when the
initial state is given by
\begin{equation}
\mathcal{N}_{p}\left( \left\vert \alpha (0)\right\rangle
_{1}+\left\vert -\alpha (0)\right\rangle _{1}\right) \left\vert
\alpha (0)\right\rangle _{2}\left\vert \left\{ 0\right\}
\right\rangle _{R}.  \label{19}
\end{equation}%
This example is the analogous of the \textquotedblleft
Schr\"{o}dinger cat state\textquotedblright\ studied in
Ref.\cite{brune96} in the context of unimodal cavity QED, in the
microwave domain, to unravel the role of the environment in the
transition from the quantum to classical dynamics. For calculating
the corresponding evolution of the SCS of Eq.(\ref{19}), it is
enough to consider the evolution of the initial state $\rho
_{12R}(0)$ as given by Eq.(\ref{16a}), and left $\alpha _{1}$,
$\alpha _{2}$ and $\beta _{1},$ $\beta _{2}$ assuming the values
$\pm \alpha $ to compose the complete density operator. Using
Eqs.(\ref{16}-\ref{17}) and taking the limit $D(t)\rightarrow 0$ for
zero temperature, we obtain for the evolved
mode $1$ in Eq.(\ref{16a})%
\begin{eqnarray}
\rho _{1}(t)&=&\frac{_{2}\left\langle \beta _{20}\right\vert \left.
\beta _{10}\right\rangle _{2}~~_{1}\left\langle \alpha
_{20}\right\vert \left. \alpha _{10}\right\rangle
_{1}}{_{1}\left\langle -u_{11}(t)\alpha _{20}+u_{21}(t)\beta
_{01}\right\vert \left. u_{11}(t)\alpha
_{10}+u_{12}(t)\beta_{02}\right\rangle _{1}} \nonumber \\ &\times& \left\vert
u_{11}(t)\alpha _{10}+u_{12}(t)\beta _{01}\right\rangle
_{11}\left\langle -u_{11}(t)\alpha _{20}+u_{12}(t)\beta
_{02}\right\vert . \label{18}
\end{eqnarray}%
Now, taking into account Eq.(\ref{18}), when considering the initial
state
as given by Eq.(\ref{19}), $\rho _{1}(t)$ reads%
\begin{eqnarray}
\hspace{-2.5cm}\rho _{1}(t) &=&\mathcal{N}_{e}\left\{ \left\vert \left[ u_{11}(t)+u_{12}(t)%
\right] \alpha _{10}\right\rangle _{1}\left\langle \left[ u_{11}(t)+u_{12}(t)%
\right] \alpha _{10}\right\vert \right. \nonumber \\ \hspace{-2.5cm}&+& \left. \left\vert \left[ -u_{11}(t)+u_{12}(t)%
\right] \alpha _{10}\right\rangle _{1}\left\langle \left[
-u_{11}(t)+u_{12}(t)\right] \alpha _{10}\right\vert \right. \nonumber \\
\hspace{-2.5cm}&+&\left. Z(t)\left[ \left\vert \left[ u_{11}(t)+u_{12}(t)\right]
\alpha _{10}\right\rangle _{1}\left\langle \left[
-u_{11}(t)+u_{12}(t)\right] \alpha _{10}\right\vert +h.c.\right]
\right\}
\end{eqnarray}%
where $h.c.$ means Hermitian conjugate and%
\begin{eqnarray}
\hspace{-2.5cm}Z(t) &=&\exp \left\{ -2\left\vert \alpha _{0}\right\vert ^{2}\left[
1-\left\vert u_{11}(t)\right\vert ^{2}\right. - \left. \frac{u_{12}(t)u_{11}^{\ast }(t)}{2} +\frac{u_{12}^{\ast}(t)u_{11}(t)}{2}\right] \right\}
\end{eqnarray}%
is the term responsible by decoherence. To evaluate the errors in
the SCS prepared in one of the two modes caused by the environment
and due to the
presence of the coherent state in the other mode, we calculate the fidelity $%
\digamma =_{1}\left\langle \Psi \right\vert \rho _{1}(t)\left\vert
\Psi \right\rangle _{1}$, where $\rho _{1}(t)$ and $\left\vert \Psi
\right\rangle _{1}$ is the evolved (mixed) and the prepared (ideal)
SCS, respectively, for the prepared SCS in mode $1$. After a
straightforward calculation and
rearranging the terms, we obtain%
\begin{eqnarray}
\digamma  &=&\mathcal{N}_{p}^{2}\mathcal{N}_{e}^{2}e^{-\left(
\left\vert \alpha (t)\right\vert ^{2}+\left\vert \alpha
_{0}\right\vert ^{2}\right) }\left\{ \cosh \left[ \alpha _{0}^{\ast
}\alpha (t)+\alpha _{0}\alpha ^{\ast
}(t)\right] \right.   \nonumber \\
&+&\left. \cosh \left[ \alpha _{0}^{\ast }\alpha (t)-\alpha
_{0}\alpha ^{\ast }(t)\right] \right\} [2+Z(t)+Z^{\ast }(t)],
\label{20}
\end{eqnarray}%
where the normalization factors are
$\mathcal{N}_{p}=[1+e^{-2\left\vert
\alpha _{0}\right\vert ^{2}}]^{-\frac{1}{2}}$, $\mathcal{N}%
_{e}=[2+(Z(t)e^{-2\left\vert \alpha (t)\right\vert
^{2}}+hc)]^{-\frac{1}{2}}$
for the prepared (ideal) and the evolved (mixed) states, respectively, and $%
\alpha (t)=\alpha _{0}\left[ u_{11}(t)+u_{12}(t)\right] $.

Fig.1 shows the fidelity of the prepared SCS state considering
parameters recently used in experiments with two modes of a cavity 
QED \cite{haroche01}. As usual, we have disregarded the cavity-field losses during the
considerably short time during which the atom crosses the cavity, about $%
10^{-4}s$. As expected, as the field excitation increases, the
fidelity decays faster. For $\alpha $ around unity, the fidelity
indicates a large time, as compared with the damping time, during
which the SCS remains available for further operations.

\begin{figure}[t!]
\begin{center}
\includegraphics[{height=6.0cm,width=7.5cm}]{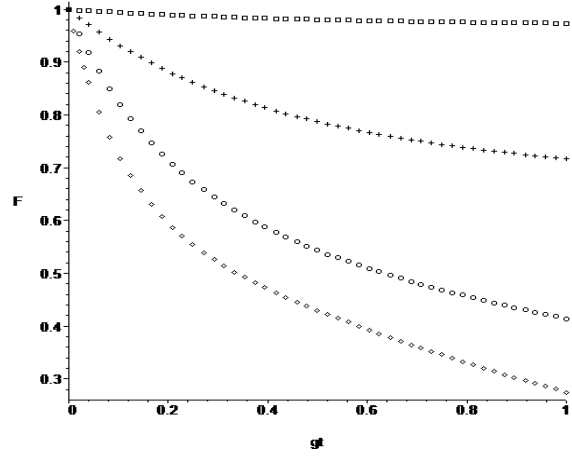}
\end{center}
\caption{Fidelity for the prepared SCS for $\protect\alpha =0.5$
(box), $1.0$ (cross), $1.5$ (circle), and $2.0$ (diamond). Here we
used the experimental values $T_{1}=$ $1$ $ms$ and $T_{2}=0.9$ $ms$
for the damping time corresponding to mode$-1$ and mode$-2$,
respectively.}
\end{figure}

In conclusion, in this paper we presented, for the first time, a
calculation of the fidelity for a superposition of coherent states
prepared in one of two modes of a cavity QED. To achieve this goal,
we solved the problem of two modes embedded in a same dissipative
environment. Different from previous studies, our problem includes
off-resonantly interaction between a two-level atom, necessary for
preparing the coherent state superposition, and one of the two
modes. Our result indicated high-fidelity of the prepared
superposition of coherent states when its mean photon number is on
order of unity.

\bigskip We thank the VPG/Vice-Reitoria de P\'{o}s-Gradua\c{c}\~{a}o e
Pesquisa-UCG (NGA), CAPES(IPQ, WBC), and CNPq (NGA), Brazilian
Financial Agencies, for the partial support.

\end{document}